\documentclass[letterpaper,twocolumn,10pt]{article}

\usepackage[utf8]{inputenc}
\usepackage{graphicx}
\usepackage{enumitem}
\usepackage{usenix2022_SOUPS}
\usepackage{tabularx}
\newcolumntype{Y}{>{\centering\arraybackslash}X}
\usepackage{xurl}

\begin{document}


\title{\Large \bf Usability Assessment of the OnlyKey Hardware Two-Factor Authentication Key Among Low Vision or Blind Users}

\def\plainauthor{Aziz Zeidieh, Filipo Sharevski}

\author{
{\rm Aziz Zeidieh}\\
DePaul University
\and
\rm Filipo Sharevski\\
DePaul University
}

\maketitle
\thecopyright

\begin{abstract}
Hardware security keys undoubtedly have advantage for users as ``usability'' pain is trivial compared to the maximum ``security'' gain in authentication. Naturally, the hardware factor in the authentication received a widespread adoption amongst average users, as it is ergonomically less demanding than phone texts or authentication prompts. This ergonomic advantage in particular is essential for users who are blind or low vision, as their interaction with a phone is impractical. However, the ``usability'' for low vision or blind users'' pain might be much higher than an average well-bodied user for the same ``security'' gain. In an effort to learn more we conducted a usability assessment with ten low vision or blind users setting up the OnlyKey two-factor authentication key. First, the setup process was insurmountable for more than half of the participants, resulting in a situation where the hardware key was abandoned. Secondly, the lack of tactile orientation led participants to consider it as both impractical, and prone to difficulties locating or loosing it. We discuss the implications of our findings for future improvements in usable authentication for visually impaired users.  
\end{abstract}

\section{Introduction}
Past research in the area of hardware security key usability \cite{reynolds_tale_2018, das_qualitative_2018} has primarily centered around non-disabled users. Narrowing down the scope of research to people who are low vision or blind, the available work on the topic of hardware security key accessibility and usability is quite paltry. This is  surprising, given the existence of frameworks and agendas for designing inclusive security and privacy mechanisms \cite{Wang, Wolf}. Barbosa et al. designed a password manager application for people who are low vision and blind \cite{barbosa_unipass_2016}. Azenkot et al. developed PassChords, a visual authentication method for touch surfaces that is robust to aural and visual eavesdropping \cite{Azenkot}. Another work along this lines is BraillePassword by Alnfiai et al., a web based authentication mechanism for blind users \cite{Alnfiai}. These works are important in addressing the needs of the visually impaired users but none address the case of second factor of authentication when it comes to usability of external hardware keys.

\section{Research Study}

\begin{figure}[h]
    \centering
    \includegraphics[width=4cm]{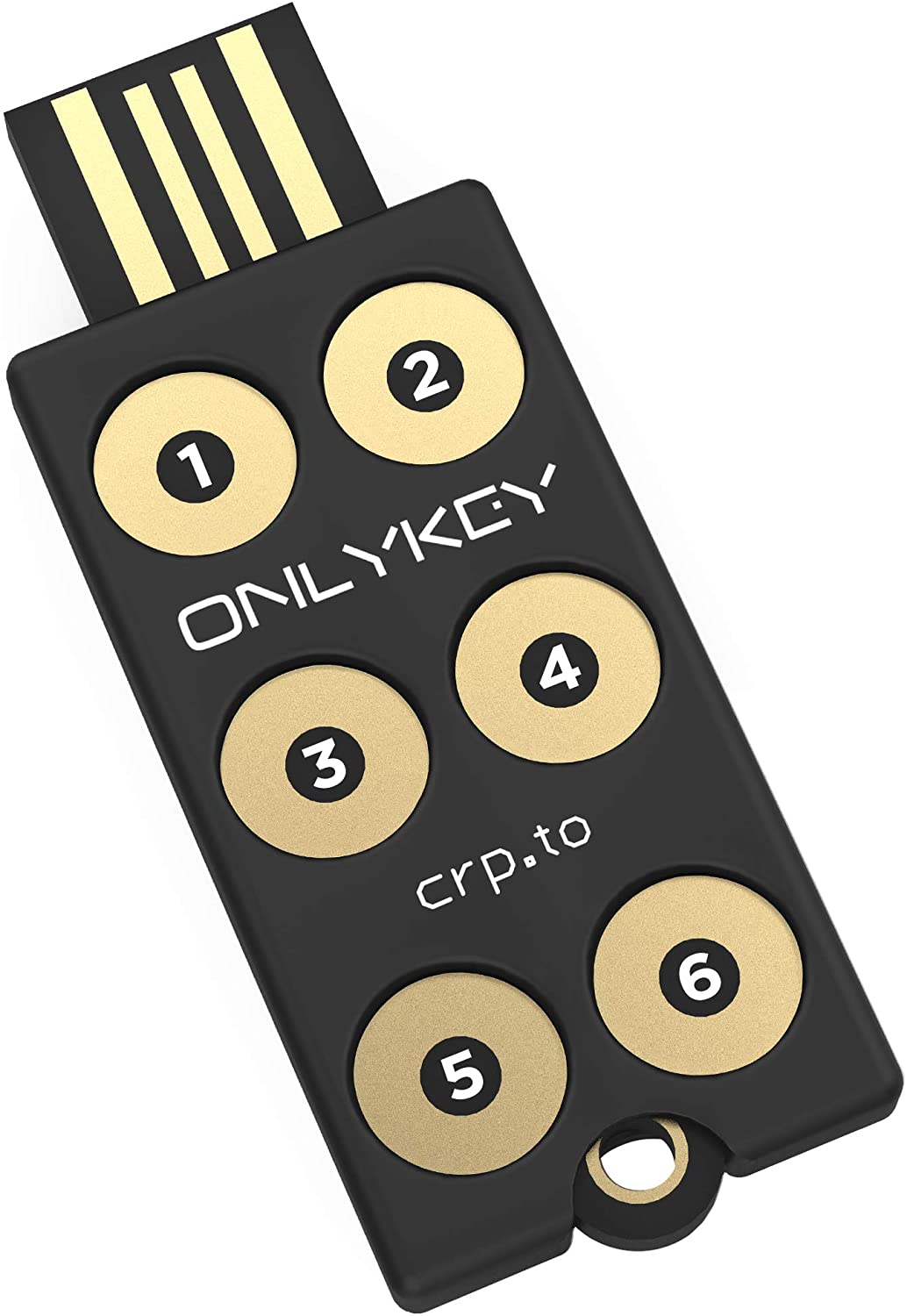}
    \caption{An image of the OnlyKey hardware security key}
    \label{fig:onlykey}
\end{figure}

This research evaluates the accessibility and usability of OnlyKey, a hardware security key that offers multi-factor authentication, and password management functionality show in in Figure \ref{fig:onlykey}. OnlyKey is a USB device with the form factor of a common flash drive. It comes fitted in removable silicon case. On one side, OnlyKey has six solid state buttons, numbered one thru six resembling a six dot braille cell, and on the reverse side there is an LED indicator light. While the arrangement of the physical solid state buttons is identical to a six dot braille cell, the numbering does not conform to the braille cell dot numbers. The male USB A plug on OnlyKey has the PCB exposed instead of encasing it in a metal structure as commonly seen in USB devices outside of the hardware security industry. The top edge of OnlyKey has a key ring hole, and comes with a quick release connector that can be used to connect OnlyKey to the user's keys. 

\subsection{Sampling}
Our research will exclusively focus on the accessibility and usability of the OnlyKey hardware security key, by users who are low vision or blind. OnlyKey is a hardware security key that offers password manager and multi factor authentication functionality, similar to the YubiKey by Yubico \cite{noauthor_yubico_nodate, noauthor_cryptotrust_nodate, noauthor_onlykey_nodate}.
 We got approval from our Institutional Review Board (IRB) to conduct semi-structured interviews with a sample of ten visually impaired users from the United States. All participants were 18 years or older. We used snowball sampling technique as the initial couple of participants were asked to recommend another participant with visual impairments for the study. 
The interviews were conducted over Zoom, as in-person presence was restricted by the university policy. The interviews lasted between 60 and 90 minutes, and users were shipped the OnlyKey hardware key beforehand. Compensation for participation in the research study was an OnlyKey hardware security key and a \$25 Amazon gift card.

\subsection{Data Collection}

The data collection for this research was primarily made up of two parts. The first part was the hands-on portion of the interview, where participants were directed to familiarize themselves with the OnlyKey, and set it up using the OnlyKey Quick Setup option. 
The second part of the interview script consisted of thirteen questions, ten were Likert scale questions adapted from the Accessible Usability Scale (AUS) \cite{virani_accessible_nodate}, whereas the final three questions, were open-ended questions, designed to gauge participants attitude toward the accessibility and usability of OnlyKey, given their experience with it in the hands-on portion of the interview. 

\section{Results}

Ten people who are low vision or blind agreed to participate in this research study as interviewees. Participants will be herein referred to as \textbf{P1} thru \textbf{P10}. Eight of the participants identified as male  while two identified as female. Three participants identified themselves as having low vision with limited visual acuity (B3), two participants identified as being totally blind and cannot see any lights or shapes (B1), two as being blind and having visual perception to see only lights and shapes (B2), two participants who stated they had low vision with high visual acuity (B3), and one participant identified as being low vision but not legally blind (B4). For the remainder of this paper, participants will be categorized into one of four visual classifications -- B1 through B4 -- based off of their response to the visual perception question in the demographics section of the interview. These four classifications are used by the United States Association of Blind Athletes (USABA) \cite{noauthor_visual_nodate}. 

Participants' responses to the AUS were calculated and included in Table \ref{tab:AUC} next to their visual classification level. For the positively worded statements (questions 1, 3, 5, 7, and 9), the score contribution is identified by taking the scale position and subtracting 1. Then, multiplying the resulting number by 2.5. For the negatively worded statements (questions 2, 4, 6, 8, and 10), the score contribution is 5 minus the scale position. Then, multiplying the resulting number by 2.5. \cite{virani_accessible_nodate}. The minimum possible AUS score is 0 while the maximum possible AUS score is 100. The average AUS score across all research participants shown in the table above is 46.5. 

\begin{table}[!h]
\begin{centering}
\footnotesize
\caption{AUS and Visual Classification of the Sample}
\begin{tabularx}{\linewidth}{|c|Y|Y|Y|}
\hline
    \textbf{Participant} & \textbf{Visual Classification} & \textbf{AUS Score} \\
    \hline
    P1 & B3 & 47.5 \\
    \hline
    P2 & B3 & 52.5 \\
    \hline 
    P3 & B4 & 37.5 \\
    \hline
    P4 & B3 & 35 \\
    \hline 
    P5 & B2 & 37.5 \\
    \hline
    P6 & B2 & 37.5 \\
    \hline
    P7 & B1 & 47.5 \\
    \hline
    P8 & B3 & 65 \\
    \hline
    P9 & B1 & 55 \\
    \hline
    P10 & B3 & 50 \\
    \hline
\end{tabularx}
\end{centering}
\label{tab:AUC}
\end{table}

\subsection{Initial Observations}
Research participants were asked to take time to familiarize themselves with the OnlyKey. \textbf{P6}, \textbf{P7}, \textbf{P8} and \textbf{P10} described the OnlyKey keypad as resembling a Braille cell. Shortly thereafter, all participants who's visual perception fell into the B1, B2 and B3 categories, excluding \textbf{P10}, expressed concern regarding the layout of the OnlyKey keypad, having no knowledge of which buttons are associated with which numbers. After reading the OnlyKey user manual, participant \textbf{P5} asked ``wait, which button is which though?'' Participants were also thrown off by the atypical USB A plug on OnlyKey. While this style of USB A plug is common with security keys like the YubiKey, all participants in this research had no prior experience with any security keys. Some participants would later go on to plug the OnlyKey into the computer upside down. 
\textbf{P4} initially identified the six buttons on OnlyKey and referred to them as a ``netting woven into the device.'' \textbf{P4}, \textbf{P5} and \textbf{P7} used an app called \textit{Be My Eyes} \cite{noauthor_be_nodate}.

\textbf{P10} did not use assistive technologies during the entire span of the research study. They were unsure if the keys on OnlyKey were ``buttons'' but they stated  ``I guess that will be something I'll find out in setup?'' \textbf{P6} employed the assistance of their sighted relative to familiarize themselves with the OnlyKey keypad layout. 
They asked their relative what the orientation of the keypad was, and where each number key was. \textbf{P5}, \textbf{P7} and \textbf{P8} attempted to locate a written description of the OnlyKey layout in the OnlyKey manual and on Google, they were unable to find anything to help address this inquiry. 

 \subsection{OnlyKey Quick Setup}
\textbf{P4}, \textbf{P5}, \textbf{P6}, \textbf{P7} and \textbf{P8} initially found and attempted to watch the official OnlyKey video on how to set up the security key \cite{onlykey_how-_2019}. However, the video lacked narration nor spoken feedback, only text instructions, visual graphics, and a background instrumental which was of no use to the participants. \textbf{P3} and \textbf{P10} had enough usable vision to read the text on the package and successfully scan the QR code which would take them to the OnlyKey ``start'' page.
\textbf{P1}, \textbf{P2}, \textbf{P4}, \textbf{P5}, \textbf{P6}, \textbf{P7} and \textbf{P8} used an app called \textit{Seeing AI}, to read the text off of the OnlyKey packaging. \textit{Seeing AI} is an app by Microsoft that enables people who are low vision or blind to see the world around them with artificial intelligence . 
\cite{noauthor_seeing_nodate}.
A photo of the OnlyKey retail packaging is in the appendix. 

While \textbf{P1}, \textbf{P2}, and \textbf{P4} were able to successfully get the URL of the OnlyKey online start page, \textbf{P5}, \textbf{P6}, and \textbf{P7} struggled with \textit{Seeing AI} to get this information, abandoning the packaging as a source of information. 
\textbf{P8} had some usable vision and noticed the QR code on the back of the OnlyKey package and attempted to use the product identification function of \textit{Seeing AI} to scan the QR code. \textit{Seeing AI} does not recognize QR codes in the product identification mode.  

Those who did not rely on the packaging as a source of information used Google or Bing. All participants ultimately found the correct instructions to follow for the OnlyKey Quick Setup process. \textbf{P1}, \textbf{P3}, \textbf{P5} and \textbf{P10} were the only participants who were able to successfully set up the OnlyKey and know the PIN configured on the device after the setup process. 
\textbf{P2}, \textbf{P4}, \textbf{P6}, \textbf{P7}, \textbf{P8} and \textbf{P9} were technically able to set up the OnlyKey, however, they did not know the PIN that was configured on the OnlyKey. This resulted in them being locked out of OnlyKey requiring them to reset it to factory defaults after the interview.

The OnlyKey Quick Setup method requires that a user plug the OnlyKey into their computer, open up a text editor, (NotePad on Windows or TextEdit on macOS), then press and hold the 3 key on the OnlyKey keypad for five or more seconds then release. With the OnlyKey acting as a keyboard connected to the computer, it types out text at the point of the cursor.
At this point of this process, the text ``you have 20 seconds'' is printed out referring to the time the user has to choose between a custom PIN, or a randomly generated PIN, with randomly generated being the default if no response is recognized after 20 seconds. 

\textbf{P2}, \textbf{P4}, \textbf{P6}, \textbf{P7}, \textbf{P8} and \textbf{P9} all used a screen reader. This is important because those who tried interacting with their computer while the OnlyKey setup process was in progress shifted the focus of the cursor resulting in text being typed by OnlyKey to go elsewhere. 
For example, \textbf{P4} could hear their screen reader echoing out typed text as the OnlyKey printed out the instructions, however, the printed text was intelligible given the speed at which the OnlyKey was typing. \textbf{P4} tried to read through the typed out text with their screen reader by using the arrow keys to navigate the text, which also shifted the position of the cursor while the OnlyKey was still printing out instructions. This resulted in the OnlyKey instructions being printed out of order and in a jumble which would ultimately lead to the OnlyKey getting setup with the PIN in the text file that the participant was unable to discern.

\textbf{P6} and \textbf{P7} were pressing modifier keys associated with functions of the screen reader to read through the document, while OnlyKey was still printing out text. This resulted in shifting the focus of the cursor outside of the text document all together. In \textbf{P6}'s case, the focus of the cursor shifted outside the text editor which resulted in output from OnlyKey being entered in other open applications. \textbf{P10} was able to see the screen and did not interact with the computer while OnlyKey went through the Quick Setup process of typing out text in NotePad. While the OnlyKey was typing out text, \textbf{P10} was reading through the manual and missed the 20 second prompt asking if they want to choose a custom PIN or have one randomly generated. The OnlyKey by default opted to generate random PINs, printed them out in the text document, and finalized the setup process while P10 was reading the user manual.

\textbf{P4}, \textbf{P6}, \textbf{P7} and \textbf{P8} realized that the setup had not gone as planned and deleted whatever text was in the text document, unplugged the OnlyKey, plugged it back in, and proceeded with the OnlyKey Quick Setup steps once more. They were confused as to why this did not work again as planned expressing frustration with \textbf{P6} saying: ``what! It's not doing it anymore! I'm doing the same thing!'' At this point, the participant's OnlyKey was setup and configured with a PIN which means the OnlyKey Quick Setup would not be available anymore. After participants exhibited signs of frustration and stress over this process, the researcher intervened to notify the participant that the OnlyKey had been setup at this point, and explained how that came to be. 
\textbf{P1}, \textbf{P3} and \textbf{P10} were able to set up the OnlyKey following the OnlyKey Quick Setup instructions 
outlined in the manual. \textbf{P5} assumed that only one key on the OnlyKey keypad would result in text output as part of the OnlyKey Quick setup process, so their initial approach was to press and hold random keys for five or more seconds until they got the expected result. In \textbf{P5}'s efforts of finding the 3 key, they eventually were able to get the OnlyKey to output instructions and PINs. However, instead of the 3 key, \textbf{P5} had randomly chose the 2 key. Pressing and holding the 2 key for five or more seconds then releasing results in OnlyKey going through the Quick Setup process with ``random PINs generation'' as the behavior instead of offering the prompt during what would have been the traditional Quick Setup through the 3 key. 

After \textbf{P5} reviewed the content in the text document, they were able to get the PIN they needed to unlock the OnlyKey, but at this point, \textbf{P5} still did not know the exact layout of the OnlyKey keypad, and this is when they called a sighted volunteer through \textit{Be My Eyes} to ask for a verbal description and orientation of the OnlyKey keypad \cite{noauthor_be_nodate}. At this point, all participants, regardless of ability to set up OnlyKey, were debriefed prior to proceeding to the post-experiment survey. All participants were made aware of the alternative method of setting up OnlyKey, which involved an application. Participants who were unable to set OnlyKey up were provided details on what issues they encountered. 

\subsection{Post Experiment}

As part of the post-experiment survey, participants were asked three open-ended questions about the OnlyKey. The first question asked what they like about OnlyKey from an accessibility standpoint, while the second asked what they disliked.
The third and final open-ended question asked for any questions, comments, concerns, or complaints the participant may have had regarding OnlyKey, both for accessibility and in general. 

Participants expressed interest in what the OnlyKey had to offer in terms of features and functionality. OnlyKey's familiar form factor was a point brought up by some participants as a positive aspect. Its also important to note that the form factor and design of OnlyKey was determined to be a negative aspect of the device by other participants. 
Participant attitudes towards the OnlyKey throughout the hands-on experiment became almost predictable after the researchers completed a few interviews with prior subjects. After participants were introduced to the OnlyKey, they expressed a sense of excitement and curiosity, however, as the hands-on experiment progressed, their excitement dwindled, and their curiosity would morph into frustration and confusion. 

A primary aspect of the OnlyKey that caused this predictable transformation mid-interview can be attributed to the physical design of the OnlyKey, more specifically, the absence of device feedback that can be interpreted by someone who is low vision or blind, similar to tactile or auditory feedback. 
Since OnlyKey has solid state buttons, the only feedback is visual, through the single LED indicator light on the back of OnlyKey. 

Participants noted this flaw in their responses to open-ended questionswith \textbf{P3} saying ``I don’t like the setup process. It would be nice if there
was non-visual feedback when clicking the buttons, not just the light.'' \textbf{P10} was able to eloquently summarize the majority of complaints brought up by prior participants in their response ``I disliked that the numbers did not correspond with the braille layout. 
It took me a moment to realize the buttons were not ‘clicky'' buttons, and I did not like how it only gave me 20 seconds.''

Another notable complaint shared by participants was the lack of detail in the instructions for a user who is low vision or blind. The instructions provided no verbal description of the OnlyKey's keypad layout, and the official OnlyKey instructional videos had no spoken feedback, only visuals and instrumental. 

All participant responses to the open-ended questions can be found in the appendix. Qualitative findings for this research study were analyzed manually by the researchers. Code books were not used in the analysis of these findings.   

\section{Discussion and Conclusion}
The objective of this study was to evaluate the accessibility of setting up the OnlyKey by people who are low vision or blind. A majority of participants were unsuccessful in the setup of OnlyKey with 60\% of participants ultimately being unable to achieve this task.
Of the four participants who were able to set the OnlyKey up successfully, three had usable vision, and the fourth only had perception of light. The participant who was blind and can only see lights and shapes relied on the help of a sighted volunteer from \textit{Be My Eyes}
\cite{noauthor_be_nodate}. 
Our usability assessment of OnlyKey as a Current-Of-The-Shelf (COTS) hardware authentication key strongly indicates that the design fails to be inclusive for the usability of the visually impaired users.
This is problematic as these users are deprived of the opportunity to benefit from most if not all functionalities provided by OnlyKey. No hardware security key on the market as of the time of this writing offers all the functionality and versatility that OnlyKey offers, which forces this population to compromise maximum security and opt for an inferior security key.  

We acknowledge that we had a limited number of participants in this hands-on study in comparison with the general population of individuals who are low vision or blind. We also are aware that the 60 to 90 minutes should have not been enough for the participants to familiarize with the OnlyKey properly. 
Another notable limitation was the choice of setup option of the OnlyKey. This research evaluated the accessibility of OnlyKey using the OnlyKey Quick Setup option and did not explore the OnlyKey desktop software. Findings gathered by this research were not interpreted based on the technical aptitude of participants. Result interpretation was focused on participant's visual perception without regard for age, gender, education, prior knowledge of hardware security keys, or proficiency with computers.

\section*{Acknowledgements}

A special thank you to Amy Gabre, Jackie Jackson, Jeni Shaum, Sue Dalton, and Wendy Brusich.

\newpage

\bibliographystyle{plain}
\bibliography{usenix2022_SOUPS}

\newpage

\appendix
\section*{Appendix}
\section{Hands-on Tasks}

In the hands-on portion of the interview, participants were asked to perform the below tasks. 

\begin{enumerate}[label=\textbf{T-\arabic*:}]
\item This is the OnlyKey hardware security key by CryptoTrust. Please take 10 minutes to familiarize yourself with it using any resource at your disposal.
\item Please describe the physical characteristics of the OnlyKey?
\item Please use the OnlyKey Quick Setup feature to perform the initial setup of this OnlyKey.
\end{enumerate}

\section{Accessible Usability Scale Questions}

There are five possible responses to any given Accessible Usability Scale question: 
\begin{enumerate}
\item \textbf{Strongly Disagree}
\item \textbf{Disagree}
\item \textbf{Neutral}
\item \textbf{Agree}
\item \textbf{Strongly Agree}
\end{enumerate}

Participants were asked to respond to the below questions adapted from the Accessible Usability Scale, (AUS).

\begin{enumerate}[label=\textbf{AUS-\arabic*:}]
\item I would like to use the OnlyKey frequently, if I had a reason to. 
\item I found the OnlyKey unnecessarily complex.
\item I thought the OnlyKey was easy to use.
\item I think that I would need the support of another person to use all of the features of the OnlyKey.
\item I found the various functions of the OnlyKey made sense and were compatible with my assistive technology.
\item I thought there was too much inconsistency in how the OnlyKey worked.
\item I would imagine that most people with my assistive technology would learn to use the OnlyKey quickly.
\item I found the OnlyKey very cumbersome or awkward to use.
\item I felt very confident using the OnlyKey.
\item I needed to familiarize myself with the OnlyKey before I could use it effectively.
\end{enumerate}

\section{Open-ended Questions}

The three open-ended questions that followed were designed to gauge the participant's overall experience with the OnlyKey. These open-ended questions were written by the researchers and are designed primarily to gain a deeper understanding of individual experiences that could not be conveyed in responses to attitude Likert scale questions. 

\begin{enumerate}[label=\textbf{Q-\arabic*:}]
\item What did you like about the OnlyKey from an accessibility and usability standpoint?
\item What did you dislike about the OnlyKey from an accessibility and usability standpoint?
\item Thinking of your experience with the OnlyKey hardware security key, what are some final comments, questions, concerns, and/or complaints you have regarding accessibility and usability of this product?
\end{enumerate}

\section{Open-ended Questikon Responses}

Participant responses to the open-ended question ``What did you like about the OnlyKey from an accessibility and usability standpoint?''

\begin{itemize}
    \item \textbf{P4:} ``I liked that it was a flash drive and that it was easy to familiarize myself with how to plug it in and kind of how to use it'' 
    \item \textbf{P5:} ``I liked that it worked in notepad so you did not have to use the app and it has tactile feel so you know where the touch sensors are.''
    \item \textbf{P6:} ``I liked that the user guide was easily found when I googled it, and was completely accessible.''
    \item \textbf{P10:} ``I liked the size of it, the packaging, the QR code and the directions were quite easy.''
\end{itemize}

Participant responses to the open-ended question ``What did you dislike about the OnlyKey from an accessibility and usability standpoint?''
\begin{itemize}
    \item \textbf{P1:} ``Initial setup with OnlyKey Quick Setup should be more clarified for users who utilize a screen reader.''
    \item \textbf{P5:} ``No audio feedback within the video, and also in terms of typing in the pass-code on the OnlyKey. It gives me flashes, but if I was totally blind that would not help me.''
    \item \textbf{P6:} ``I disliked that there was no tactile feedback on the OnlyKey when clicking the buttons. I disliked that the setup was convoluted and did not work well and was typing it in manually, it was very weird it was typing it out like a person was typing it out. I did not like the video.''
    \item \textbf{P7:} ``It's not very usable, it does not seem to be accessible from the get-go. It could be impractical from a novice with the quick setup in mind.''
    \item \textbf{P8:} ``The spitting out text is very nice, but the way it did it made it very easy to mess the process up, which I did, so that's something to consider. It's nice that there are physical buttons, but they don't give much tactile feedback, so that kind of sucks.''
    \item \textbf{P10:} ``I disliked that the numbers did not correspond with the braille layout. It took me a moment to realize the buttons were not `clicky' buttons, and I did not like how it only gave me 20 seconds.'' 
\end{itemize}

Participant responses to the open-ended question ``What questions, comments, concerns, complaints or feedback did you have regarding OnlyKey from an accessibility and usability standpoint, and in general?''  
\begin{itemize}
    \item \textbf{P3:} ``I definitely did not like the setup process, it's not straightforward. I would like there to be a more direct way to set it up. I think it's great, I really do like it, but I don't like the setup process. It would be nice if there was non-visual feedback when clicking the buttons, not just the light."
    \item \textbf{P4:} ``It being a small device, it is really easy to lose, especially for someone who is totally blind. Maybe have an interface or something where it says what button you are pressing.''
    \item \textbf{P5:} ``Obviously there's the documentation not having information on the orientation and layout of buttons. It did not tell you what mode it is in, maybe have better instructions.''
    \item \textbf{P6:} ``I had a sense of excitement acquainting myself with the onlykey and reading the user guide. Once i started installing it, I quickly got frustrated. it sounded easy, plug in, press button, hold button, but it didn't do it. But it sounds like a product I'd like to use. With prior experience, I know setting it up would be the hard part, but once setup I would think it's easy to use.''
    \item \textbf{P7:} ``Have a section that says 'you push the three button, give an approximation of how long you wait', so that you understand what's going on and when.``
    \item \textbf{P10:} ``I wish there was some more directions with the buttons, I just think that someone who is fully blind, how would they know which button is which.''
\end{itemize}

\section{OnlyKey Retail Packaging}

\begin{figure}[htp]
    \centering
    \includegraphics[width=4cm]{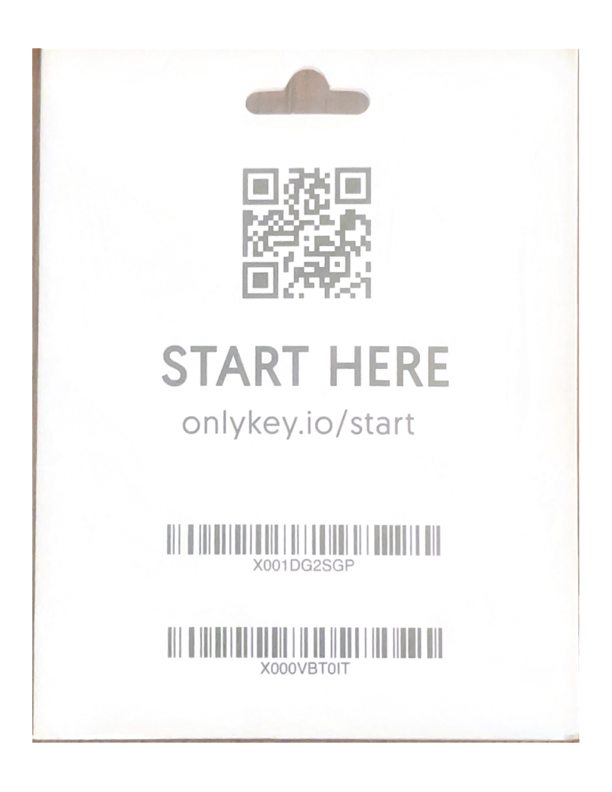}
    \caption{An image of the OnlyKey retail packaging}
    \label{fig:onlykey-package}
\end{figure}
The OnlyKey package features a QR code, the text ``START HERE'', followed by the text, ``onlykey.io/start''. At the bottom of the package, there is a barcode. 

\end{document}